# Optically Active and Nonlinear Chiral Metamaterials

Sean P. Rodrigues[1], Preston Cunha[2], Kaushik Kudtarkar[2], Ercan M. Dede[1], Shoufeng Lan[2,3]*

1 Toyota Research Institute of North America, Ann Arbor, MI 48105, USA

2 Department of Mechanical Engineering, Texas A&M University, College Station, TX 77840, USA

3 Department of Materials Science and Engineering, Texas A&M University, College Station, TX 77840, USA

Corresponding Author e-mail address: shoufeng@tamu.edu

## Abstract

Advanced photonic nanostructures have enabled the maximization of synthetic chiroptic activities. The unique structuring of these building blocks has empowered chiral selective interactions with electromagnetic waves in plasmonic structures and dielectric media. Given the repertoire of optimized chiral surfaces in the literature and the ubiquity of chirality in the organic realm, the natural direction to consider is the operation of these devices in larger optical systems much like their chiral organic counterparts. In this review, we recapitulate advances in active and nonlinear chiral metamaterials. Many of the results, such as the magneto-chiral anisotropy and third-harmonic Rayleigh scattering optical activity, are relatively unknown members of the more conventional family of tuning methodology and nonlinear processes. We believe they are poised to play an instrumental role in designing advanced chiroptic systems for applications in biochemistry, valleytronics, spintronics, and chiral quantum optics.

# Introduction

Chiral electromagnetism, often referred to as optical activity, has two main properties that can describe the way a wave interacts with a given chiral medium. The first is optical rotatory dispersion (ORD) in which the polarization of a wave is rotated as it passes through a chiral medium. The second is circular dichroism dispersion (CD) which describes the asymmetric absorption of light upon incidence with a chiral medium.[1–3] The experimental discovery of ORD was realized by Louis Pasteur in 1856, when he noticed that pure crystalized forms of (±)lactic acid were able to rotate the polarization of a linearly polarized light wave.[4,5] To this day, this optical description of chirality is used as a characterization system to classify molecular behavior. For instance, D-glucose is naturally occurring in organic life and is capable of being broken down into an energetic form for the human body, while L-glucose is simply passed through the digestive system unutilized by the metabolism; even though both forms of glucose retain the same molecular components, their conformation (not configuration) is slightly different. This alternate conformation arises from the tetrahedral nature of carbon; the bonding angles of these atoms induce alternate forms even if the atoms are connected in the same delineation in both molecules. This naturally occurring asymmetry requires extreme care in the asymmetric synthesis of chemicals and pharmaceuticals. Research in this field has led to Nobel prizes in 2001 and 2021.[6–8] However, the ability to measure these optical signals in small samples of organic molecules is often difficult due to their optical activity appearing in the UV region and the variation being often on the order of millidegrees. While measurement components like lasers, photodetectors and the like have improved, the critical functionality of how to measure chiral optical activity from organic materials has not changed much since the inception of these measurements systems. For instance, measurements of the first systematic observation of cotton effects stemming from protein conformations in 1966 rely on polarimeters similar to what we use today. [9–12]

This introduction, largely descriptive of the realm outside of chiral metamaterials, gives insight into the necessity that has driven the exploration of artificially produced optical activity. Some of these important objectives include: can large chiral optical activity be synthetically achieved; can the sensitivity of optically active signals be increased; and how can chiral-selective behavior be implemented in electromagnetic applications like its operative counterpart in biology? These directives have been extensively researched over the past several years, leading to significant advancements through the utilization of chiral metamaterials. Here, we provide demonstrative references for the above objectives using chiral metamaterials, respectively, for optical activity,[13–21] chiral-selectivity,[22–24] and applications.[25–27] In this review, the term "metamaterials" describes the structuring of two or more materials with different refractive indices to form a local unit cell. This local unit cell is then periodically repeated to form a larger global material, which has an effective refractive index. Typically, optical metamaterials have a unit cell size that is on the order or smaller than the wavelength of the radiation of interest.[28] In particular, chiral metamaterials utilize structural asymmetry at the local level to induce chiroptic responses. Chiral metamaterials created from plasmonic materials such as gold, silver, and titanium have been utilized to create strong chiral resonances in the visible and near-infrared regime. Several papers describe how to create strong chirally resonant optical devices by understanding the interplay of the polarization of incident light with the three-dimensional design of the optical structure.[29] While the advancement of nano-optics artificial chirality has grown rampantly in the past several years, it should be noted that the use of artificial materials to manipulate the polarization of electromagnetic waves dates back to the 1890's, when Bose observed that jute rope could induce a rotation on the polarization of a millimeter wave.[30]

In this review, the authors highlight recent progress on active and nonlinear chiroptical metamaterials. In particular, we refer to "active" as the dynamic response of a metamaterial to an external stimulus, which modifies the output of the light that has been incident on the metamaterial. The goal of this review is to focus on the temporal modulation of these materials, rather than the fixed tailorable response. In the past five years, there has been a surge of research on the modification of the local or effective refractive index of media, thus enabling the research of a diverse set of switchable or temporally-tailorable optical properties. Many paths of research that are related, but out of the scope of this review, include a review on multidimensional nanoscopic chiral optics[31], a review on chiral quantum optics,[32] and reviews on active metamaterial structures in general.[33,34]

Before discussing the role of active photonics in chiral artificial media, first a review of definitions and terms will be presented. Circular dichroism is defined as the difference in absorption between left and right circular polarized waves, $CD = A_{RCP} - A_{LCP}$. However, as many optical metamaterials often demonstrate very low values for reflectance, values of transmission are often used in place of absorption. It should be noted, that circular dichroism dispersion is intimately related to optical rotation dispersion via the Kramers Kronig relations.[35,36] Also important to note, a robust way to understand the Jones matrix of a material is to understand the polarization and intensity of light prior and after incidence onto the structure; to this end, Stokes polarimetry can provide a detailed analysis of the light and thus its interaction with the material.[37] Given the Stokes vectors, the handedness of the input and output waves can be depicted on a Poincare sphere for analysis. Finally, although circular dichroism is a definitive method to realize linear optical chirality, many researchers have found new and exotic ways to usurp this property through the emission of quantum dots, the use of photophores, and the implementation of nonlinear tensors in order to create extractable, larger chiral responses, many of which will be discussed in more detail below.

## 1. Active chiral metamaterials

As we understand in the previous section, the chiroptical signal generated by chiral metamaterials can largely surpass that of its organic counterparts, due to the strong coupling between electric and magnetic fields in the chiral nanostructures. With this understanding in mind, one could envision many practical applications from sensing to communications, by studying how external stimuli can modify the effective chiral features of the material, including the spectral location and magnitude.[38–40] In order to reach this step, researchers have learned how to manipulate and probe the chirality of these metamaterials actively.[41–43] To this end, the study of active photonic structures has employed many methods, including optical, electrical, mechanical means. In this next section, the review of the active structures is categorized broadly into three categories including: active chiral metamaterials with static structures for tailored coupling efficiency; and active chiral metamaterials with reconfigurable structures for control of chirality, and externally modulated chiral systems.

### 1.1. Active chiral metamaterials with static structures

For practical reasons, active chiral metamaterials with static structures are desirable. To change the coupling strength between electric and magnetic fields while keeping the structures intact, researchers use active constituting materials whose properties change with external stimuli. In the electromagnetic region, these material properties are the permittivity, permeability, or refractive index. It is possible that the electromagnetic properties of a given material are sensitive to external stimuli such as a secondary optical wave, magnetic flux, a thermal flux, etc. As chiral metamaterials are often composites of two or more materials, the effective index determined by the structuring of these materials has more opportunities to modify light than a standard bulk medium. For this reason, active chiral metamaterials can be created with static devices because the electromagnetic coupling strength that signifies the strength of chirality can be modulated with and without the stimuli. As an example, some phase-changing materials such as $VO_2$ can change from a metallic state to a dielectric state, whose electromagnetic properties are dramatically different upon the application of heat. As a result, by employing active materials as constituents in artificial chiral structures, researchers have various methods to manipulate chirality correspondingly.

**1.1.1 Optically active chiral metamaterials**

Among the variety of active stimuli to manipulate metamaterials, secondary optical sources are of great interest as they often are associated with high-speed modulation depths. The typical materials used for optical modulation are semiconductors because the conductivity changes dramatically with optical pumping at an energy higher than the bandgap (Figure 1). In the THz regime, Kanda et al. switched the chirality of their material on and off with optical pumping. This is achieved by using the distribution of photo-generated carriers in a silicon layer together in combination with a chiral gammadion-shaped metallic nano-structure.[44] The intrinsic chirality in the single-layer metallic structures is too weak to be seen by the THz waves, but the authors successfully managed this by extending the thickness with the distribution of photo-generated carriers to the thick silicon substrate. The polarization rotation angle is less than 1° at the resonant frequency. By using a bilayer conjugated metallic structure together with a thin intrinsic silicon layer, Zhou et al. further enlarged the rotation angle to tens of degrees.[45] With a judicious design, Zhang et al. introduced a metamolecule monolayer that flips the ellipticity of the chiral metamaterial.[46] Such an electromagnetic effect is much stronger than that of naturally available molecules. This is promising for creating a compact polarizer with dynamic control of the polarization of light.

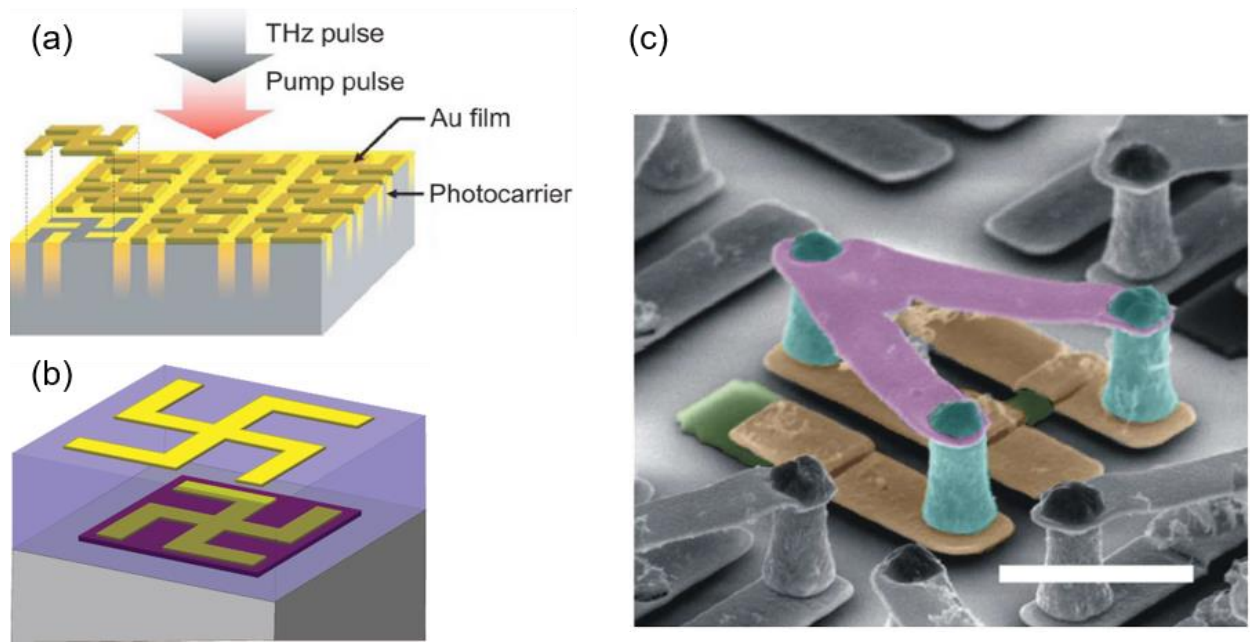

**Figure 1.** Static optically active chiral metamaterials with three-dimensionally shaped nanostructures. The photoexcited carriers in semiconductors change the electromagnetic properties dramatically and hence are used as optically active constitutive materials for chiral metamaterials. The devices contain (a) a single-layer,[44] (b) a double-layer gammadion-shaped chiral metallic structures,[45] and (c) a sophisticated V-shaped geometry bolstered by pillars.[47]

### 1.1.2 Thermally active chiral metamaterials

Compared to optical methods, thermal activation methods are usually much slower, occurring on a timescale of seconds for large bulk materials. This longer thermal time constant for activation is related to the diffusion of heat required to articulate sufficient changes in the bulk properties of the thermally active materials. However, at the nanoscale these heat induced material property changes may occur more rapidly. The most vastly available thermally modulated materials are phase change materials, whose electromagnetic properties change dramatically upon application of heat. The heat breaks the chemical bonds in the material and changes the phase of the material from crystalline to amorphous. More importantly, when the temperature cools down, the amorphous phase changes back to the more stable crystalline phase. The reversible characteristics of thermally transduced phase change materials opens a slew of real-world applications. One thing of note is that phase change materials not only can be controlled by heat, but also by other methods such as light and electricity. In this section however, we limit ourselves to the discussion of thermal methods. An example of a phase change material is $Ge_3Sb_2Te_6$ (GST-326). By applying a thermal gradient to the design of a chiral metamaterial that utilizes this phase change materials, researchers have been able to spectrally tune the resonant wavelength of the active chiral metamaterial.[48] The spectral shift is due to the dielectric constant change between amorphous and crystalline GST-326 states in the studied mid-infrared (MIR) spectral region. Subsequently, by utilizing this effect in combination with a passive chiral bias-type layer, the authors were able to switch the sign of the CD signal in the MIR, Figure 2(a). As circular dichroism is measured on a scale ranging between -1 and +1, being able to switch the overall handedness of the material, and thus the chiroptical signal from negative to positive, or vice versa, is significant in terms of modulation depth. One way to extend the resonant wavelengths to other spectral regions such as visible, NIR, or THz is to use different phase change materials.[49] Utilizing a slightly different phase change material such as $Ge_2Sb_2Te_5$ (GST), researchers have been able to reversibly tune a glass chalcogen material at nanosecond or less timescales. The phase change nanomaterial has a helical structure, as shown in Figure 2(b); the width of the rod of the helix is 15 nm, with a radius of 22.5 nm and a pitch of 45 nm. The nanostructures are created using a glancing angle deposition technique, and the metamaterial switches from exhibiting large optical activity to weak optical activity with demonstrated dynamic switching over 50,000 cycles. [50]

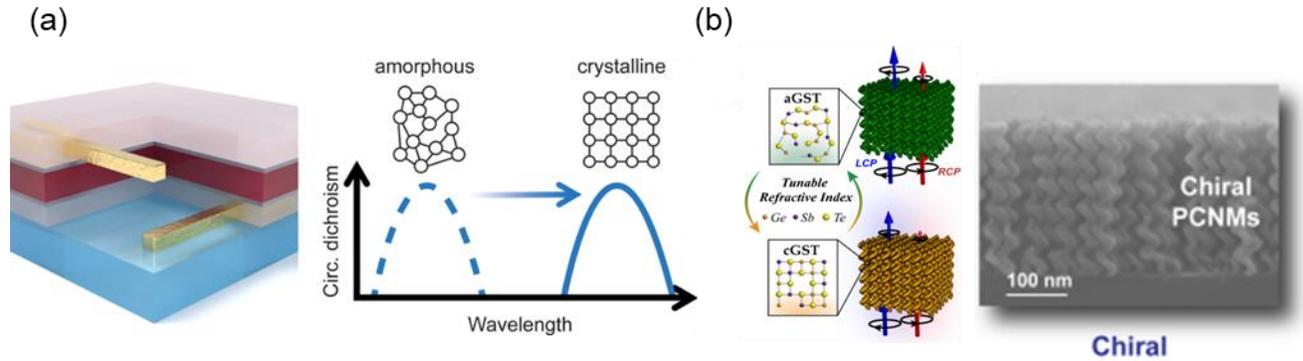

**Figure 2.** Thermally active chiral metamaterials. (a) The thermally active phase change material is GST-326, which changes from a crystalline to amorphous state when applying heat.[48] (b) A phase change nanomaterial also made of GST is capable of increasing/decreasing chirality by an external heat stimulus and shows no change in properties over 50,000 cycles.[50]

**1.1.3 Electrically active chiral metamaterials**

An alternate method for the direct tuning of chiral metamaterials is to use electrical control, which is attractive because of the compatibility with existing information processing systems, telecommunications, and so on. Unfortunately, electrical tuning of CD and optical activity (OA) is extremely challenging as the electrical doping level is hardly comparable with the total density of states in bulk materials. However, this scenario is dramatically different in two-dimensional (2D) materials. By integrating 2D graphene in a chiral metamaterial, Kim et al. tuned the transmission of a right-handed circularly polarized (RCP) light while maintaining the transmission constant for the other circular polarization. In this way, the authors were able to tune the CD by applying a gate voltage.[51] The authors further showed that they could also tune the plane of linearly polarized light while the linear polarization state maintains its linearity in the same device. The nuance here is that most conventional chiral optical systems will modify a linearly polarized wave into an elliptically polarized wave. Maintaining the linear polarization is a property that is typically observed in birefringent media. The benefit from the electric control of polarization is that the 2D material concept may lead to various applications such as a compact active polarization modulator for telecommunications and imaging devices. Similar demonstrations of graphene-embedded, THz, chiral-modulation devices have also been demonstrated.[52,53]

**1.1.4 Magnetically active chiral metamaterials**

Here, the magneto-chiroptical properties of metamaterials are investigated. Naturally, to achieve a modulation depth in the optical spectrum via magnetic tuning, the effective refractive index of the metamaterial must be reliant on a magnetically tunable material. To this end, the goal is to implement a type of interaction that is typically observed in chiral liquids known as the Cotton-Mouton effect or in more general substances, as the Faraday effect. To demonstrate this magnetic tunability, researchers introduced cobalt into the fabrication of a layered silver nanohole array, thus demonstrating a change in the modulation depth of 0.035° using this chiral metamaterial. The interplay between the two materials and the geometrical structure of the hole array enables stronger magnetic circular dichroism.[54] Beyond the first-order Faraday effect that is linearly proportional to the magnetic field, magnetism also sustains higher-order interactions with chirality stemming from their shared nature of an axial vector. One prominent example is the magneto-chiral effect, which is proportional to the inner product of the magnetic field and the wavevector.[55] This relatively unknown degree of freedom provides a new route for exotic functionalities that are generally not achievable solely by magnetism or chirality. These functionalities include enantiomeric excess in photochemical reactions and the separation of racemic mixtures. Though it is omnipresent and responds to photons, electrons, and phonons, the observed magneto-chiral anisotropy is on the order of 0.1%. By judiciously arranging gold nanoparticles on gold/cobalt magnetic multilayers as shown in Figure 3, Armelles et al. obtained a magneto-chiral response to surface plasmons with a magneto-chiral anisotropy of around 1%.[56] Very recently, Lan et al. reported the first observation of a magneto-chiral response to excitons with a record-high magneto-chiral anisotropy of ~4% in twisted van der Waals crystals.[57] Moreover, by coupling to the valleys in the electronic band structure, such metamaterials may find applications in the surging field of valleytronics.[58] This technology may also contribute to recent electron quantum metamaterials using van der Waals crystals as building blocks. [59]

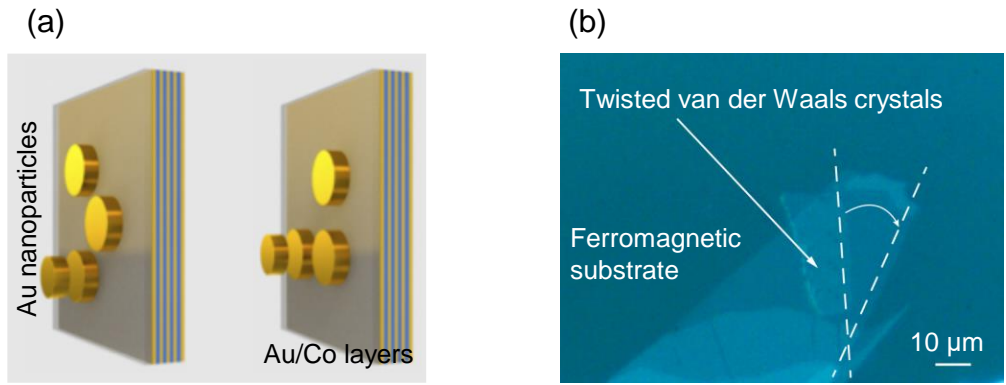

**Figure 3**. Magnetically active chiral metamaterials through the magneto-chiral effect. (a) By placing purposely arranged chiral plasmonic gold nanoparticles on gold/cobalt magnetic multilayers, the magneto-chiral anisotropy is around 1%.[56] (b) Twisted van der Waals crystals on a ferromagnetic substrate increase the magneto-chiral anisotropy to 4%.[57]

**1.2. Active chiral metamaterials with reconfigurable structures**

Intuitive thinking to tune the coupling between electric and magnetic fields and hence the chirality is to reconfigure the structures of chiral metamaterials. Imagine that as the distance between metamolecules, the thickness of layers, or the geometry of structures changes, the field distribution, as well as the coupling, must change accordingly. Indeed, with external stimuli giving rise to a strong enough force, researchers have demonstrated various active chiral metamaterials. The challenge lies in finding the right sources and materials for providing the force to introduce structural changes. Those sources can either be mechanical, chemical, or biological. Due to limited space, focus is given to the review of experimental works in this section.

**1.2.1 Mechanically reconfigurable chiral metamaterials**

Photoexcitation for chiral metamaterials with static structures can switch the handedness. On the other hand, static structures with active materials lack enantiomeric symmetry, which results in asymmetric optical activity spectra. Therefore, there is a considerable need for an active chiral metamaterial with the switching capability and symmetrical optical activity spectral shapes between the different handedness structures. Mechanically reconfigurable chiral metamaterials fulfill the need.

In this section, we review several mechanically reconfigurable structures that are actuated by various stimuli. In the THz region, Kan et al. proposed and experimentally demonstrated a micro electro mechanical systems (MEMS) chiral metamaterial facilitated by a deformable 3D chiral structure.[60–62] Later these same springs were deformed via physical movement after having the center of the spirals glued.[63] The directional switching of the pneumatic and or electrostatic actuation enabled an optical activity polarity reversal while maintaining a constant energy transmittance and spectral shape, thus achieving enantiomeric handedness switching. Also, in the Thz regime, researchers demonstrated birefringence-free tunability with pneumatically tuned helices, rather than the spirals previously discussed.

In the visible region, Kuzyk et al. used DNA as both a construction material to organize individual plasmonic nanoparticles, as well as fuel for driving the metamolecule to distinct conformational states.[64] Figure 4 provides a few more depictions of electrically controlled modifications of the reconfigurable metamaterial. Figure 4(a) depicts active control of circular dichroism by nanoelectromechanically tunable doped silicon interdigitated structures. The device depicts distinct wavelength resonance shifts, thus enabling strong shifts in CD at specific wavelengths.[65] In another example, a gold filmed metamaterial is fabricated on a substrate. By applying a voltage the metamaterial goes from a 2D thin film, to a 3D chiral nanostructure through nano-kirigami folding.[66] Finally, figure 4(c) provides mechanically folded tuning angles as well as and electrical tuning of the Fermi energy level of graphene. This paper also provides a set of videos depicting the folding animations and their respective transmission plots for both left and right handed structures.[53]

Mechanical tuning of optical activity has also been observed by strategically designing chiral metamaterials on PDMS substrates.[67] In a similar vein, researchers utilized glanced angular deposition on polystyrene nanosphere arrays to create hollow nanoshells and hollow nanovolcano film arrays; these structures are then integrated into PDMS substrates that have shown mechanical deformation to achieve active circular dichroism of the composite structure. PDMS, hydrogels and other curved plastics are highly desirable for medical sensing designs. Using the design described researchers were able to create reference tables to signify water content absorption of hydrogels.[68] This kind of deposition process, also known as shadow lithography, has been studied at length by the Zhao group for the creation of chiral nanostructures as well,[69–71] and other groups for other metamaterial structures.[72]

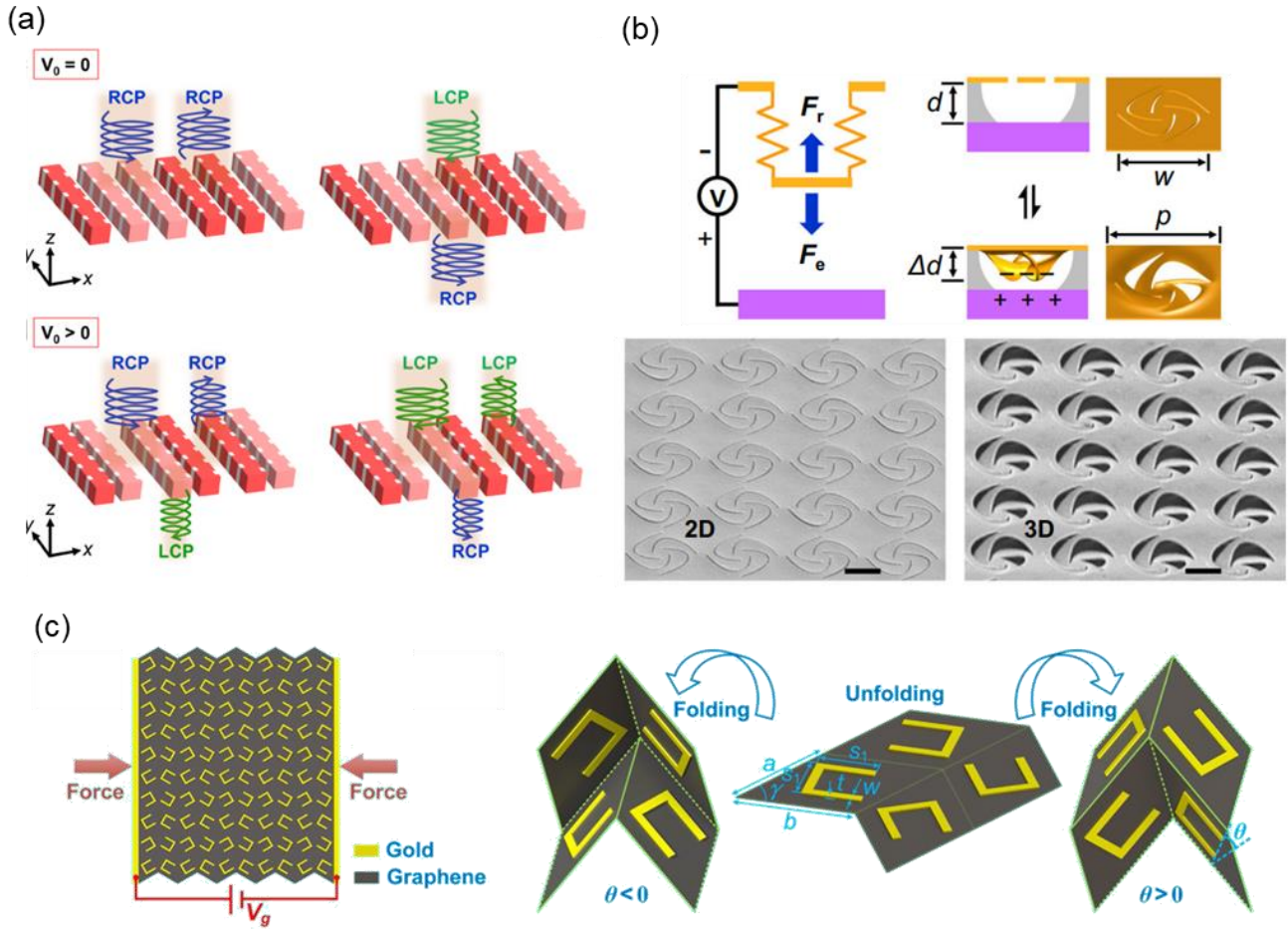

**Figure 4**. Electrically transduced mechanically reconfigurable metamaterials. (a) Interdigitated electrodes are patterned in metal. The electrodes are pulled in towards each other to instill either a right handed or left handed chiral metamaterial structure.[65] (b) A gold metamaterial membrane is bolstered by silicon dioxide pillars that sit on a silicon wafer. The silicon and the gold act as electrodes to implement the reconfigurable structure from a 2D orientation to a 3D orientation.[66] (c) Here, this G-mori type metamaterial structure shows both mechanically induced folding and electrical tuning of the Fermi level of the graphene substrate.[53]

## 1.2.2 Biochemically reconfigurable chiral metamaterials

Despite the demonstration of mechanically reconfigurable individual chiral metamolecules with DNA, active regulation of chirality with a collective motion of metamolecules, particularly in the visible spectral region, still faces significant challenges. Duan et al. demonstrated a new class of hybrid plasmonic metamolecules composed of magnesium and gold nanoparticles.[73] The chirality from such plasmonic metamolecules can be dynamically controlled by hydrogen in real-time. In this case, although gold structures did not change, we classify them as reconfigurable chiral metamaterials because part of the structures switches forms between $MgH_2$ and Mg, interchangeably. Interestingly, using a silk layer as the spacer between two twisted gold hole arrays or so-called moiré patterns, Wu et al. also demonstrated tunable chirality of chiral metamaterials in the visible region.[74] The silk spacer changes the thickness with chemical solvents and hence changes the coupling between the electric and magnetics of the chiral metamaterials. They further used this biochemically active chiral metamaterial as a sensor to detect solvent impurity of solutions; refer to Figure 5.

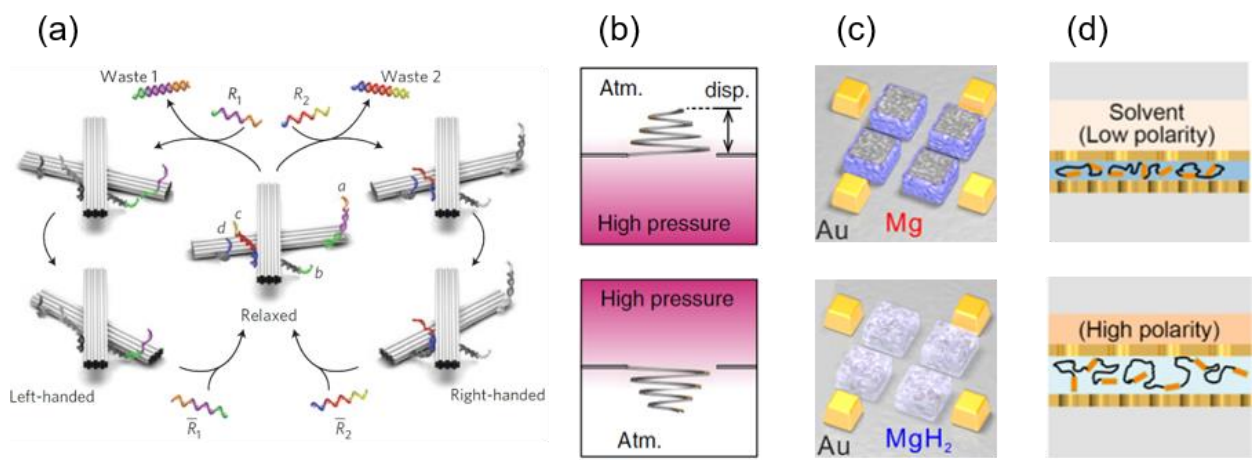

**Figure 5.** Reconfigurable chiral metamaterials. The structures are reconfigurable with (a) DNA,[64] (b) pressure,[60] (c) hydrogen,[73] (d) solvents,[74] respectively. When the structures and geometry

changes, the coupling of electric and magnetic fields, as well as chirality of the chiral metamaterials changes accordingly.

## 1.3 Modulated chiral systems

Also, worth discussing in this review are modulated chiral systems, where modulation occurs near or around the chiral metamaterial, but not directly impacting the refractive index of the metamaterial itself. Examples of this are shown in Figure 6. In one case, a liquid crystal is placed on top of an L-shaped chiral metamaterial. In this situation, the liquid crystal is able to locally alter the polarization of the incident light thereby creating an active optical system for spin-selective light absorption in the near-infrared regime.[46] In a similar vein, liquid crystals have also been used externally to realize metamaterials in the THz regime as shown in Figure 6(b).[75] In this case, the liquid crystal axis is spatially on the order of the metamaterial, thereby modifying the axial symmetry of the device. Another example of externally modulated chiral systems is that shown Figure 6(c). Here, the structure is mechanically modulated by a piezoelectric actuator. The L-shaped metamaterial is created in a GaAs membrane structure. The structure exhibits 3D chirality due to its Fabry-Perot resonances between the membrane-substrate optical paths. The optomechanical coupling is then investigated, by vibrating the membrane and imparting changes to the phase, intensity and polarization of light is controlled. The membrane oscillations are achieved by piezoelectric actuation. [76] The device has a unique feedback effect in which the polarization state of the imparted photons shifts the resonant frequency of the mechanical resonator through an optothermal spring effect. Moreover, this shift in the resonant frequency is spin-selective thanks to the directionality of the circularly polarized light.

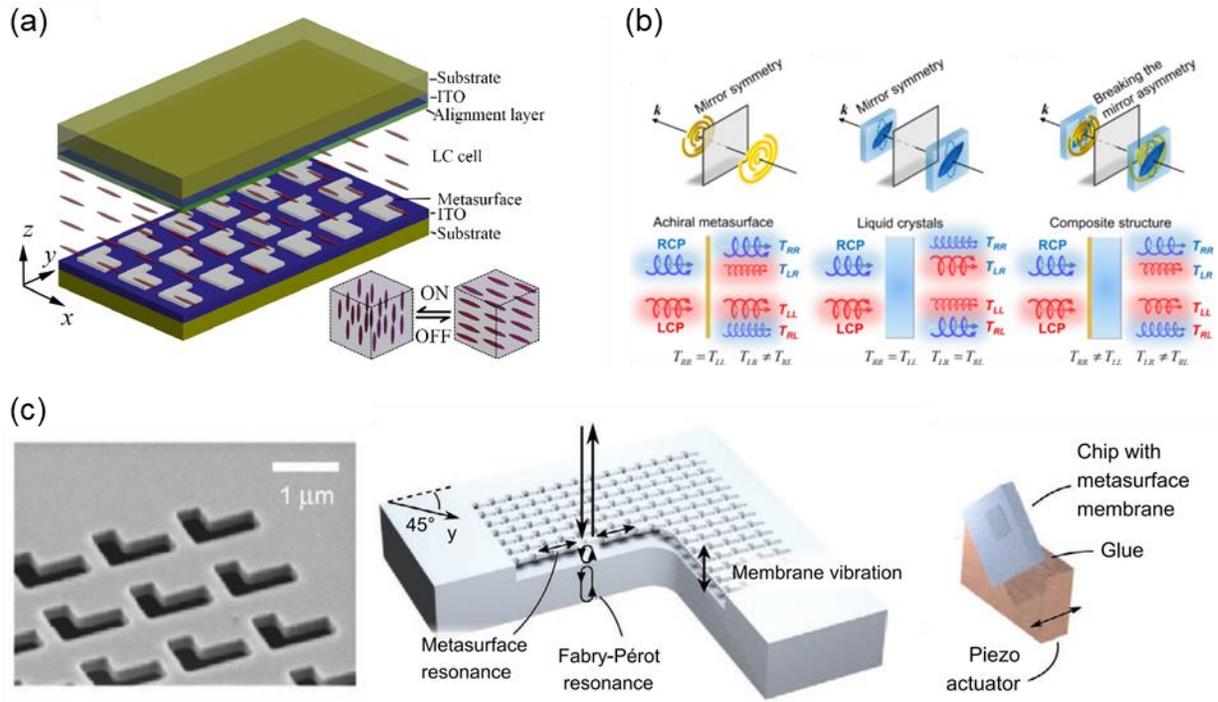

**Figure 6**. Externally modulated chiral systems. Liquid crystals placed adjacent to metamaterial structures provide optical systems (a) with spin-selective absorption in the NIR[46] and (b) that break axial symmetry in the THz regime.[75] (c) A GaAs metamaterial membrane is vibrated, while spin selective optical forces are imparted onto the membrane thanks to the handedness of the metamaterial.[76]

## 2. Nonlinear chiral metamaterials

Light-matter interactions in traditional chiral materials manifest in magnetic dipolar interactions, which is usually much weaker than the electric dipolar interactions that dominate the linear optical phenomena such as transmission, reflection, and absorption. Consequently, chiroptical properties, including CD and optical rotatory dispersion (ORD), are typically more than two-order smaller than linear optical properties such as scattering. While CD designates distinct absorptions from incident left circularly polarized (LCP) and right circularly polarize (RCP) light, ORD designates the polarization rotating ability of a chiral material. Nonlinear optical interactions such as second-harmonic generation (SHG) and sum-frequency generation (SFG), however, allow the electric dipolar interactions in media without inversion symmetry. Therefore, the nonlinear chiral responses are comparable with achiral nonlinear responses in chiral materials. For this reason, nonlinear, optical excitation tools such as SHG-CD and two-photon luminescence (TPL) CD have been developed to better identify and realize features unrecognized under linear excitation conditions.[77–80] Chiral metamaterials have proven to be powerful nonlinear optical tools in the investigation of chiroptical properties with high sensitivity.[81]

## 2.1. Second-order nonlinear chiroptical processes

The simplest but basic nonlinear optical process is SHG, in which two photons at the same frequency generate a new photon with frequency doubled in nonlinear media. The SHG is sensitive to symmetry breaking and surface conditions, therefore, it has been a critical nonlinear optical tool for investigating chiral metamaterials. Valev et al. use chiral metamaterials with the Slavic shape and its mirror-image to unambiguously observe the large distinct SHG from LCP and RCP light, Figure 7, the SHG-CD.[82] The authors attributed the large SHG-CD to the superchiral field of hotspots.[83] The authors also showed that, whereas chiroptical properties in the linear optical regime are reciprocal, in the nonlinear regime these effects can be non-reciprocal. Meanwhile, as shown in Figure 7, Rodrigues et al. introduced a twisted-arc chiral metamaterial that has an SHG-CD of ~2, the largest value on the chiral-SHG scale.[84] More interestingly, by embedding quantum dots (QDs) into the twisted-arc chiral metamaterial, the same group observed a largely enhanced TPL-CD, another second-order nonlinear chiroptical phenomenon.[23]

## 2.2. Third-order nonlinear chiroptical processes

As described previously, optical properties are generally associated with refractive index of a medium. When the refractive index changes with the intensity of the light, it generates a nonlinear optical phenomenon that typically relates to the third-order nonlinearity. In chiral metamaterials, such third-order nonlinearity leads to a novel phenomenon called nonlinear optical activity. A comprehensive study on the third-harmonic circular dichroism of bilayer metamaterials is provided by Gui et al., in which they utilized a Born-Kuhn analog to understand the chirality of nanorods arranged in C4 symmetry.[85] Ren et al. obtained a polarization rotation that depends on the intensity of incident light in a chiral metamaterial.[86] Rodrigues et al. investigated the spectral shift induced by the intensity-dependent refractive index in a structure composed of two perforated silver films.[87] With organic conjugated polymer (PFO), Chen et al. observed third-harmonic generation circular dichroism (THG-CD).[88] The authors also investigated the symmetry selection rules of THG-CD with structures of various symmetries that are coated with PFO.[89] Recently, Ohnoutek et al. obtained an optical activity by using third-harmonic Rayleigh scattering, which might provide an inspiring piece of insight for nonlinear chiroptical responses.[90] Remarkably, as the third-harmonic Rayleigh scattering is an incoherent elastic process, the resulting optical activity has no linear background, less requirement on the fundamental frequency, and is compatible with isotropic materials. The same group applied this principle to study semiconductor nanohelices that allowed for the chiroptical characterization of sample volumes as small as $10^{-5}\mu l$. [91]

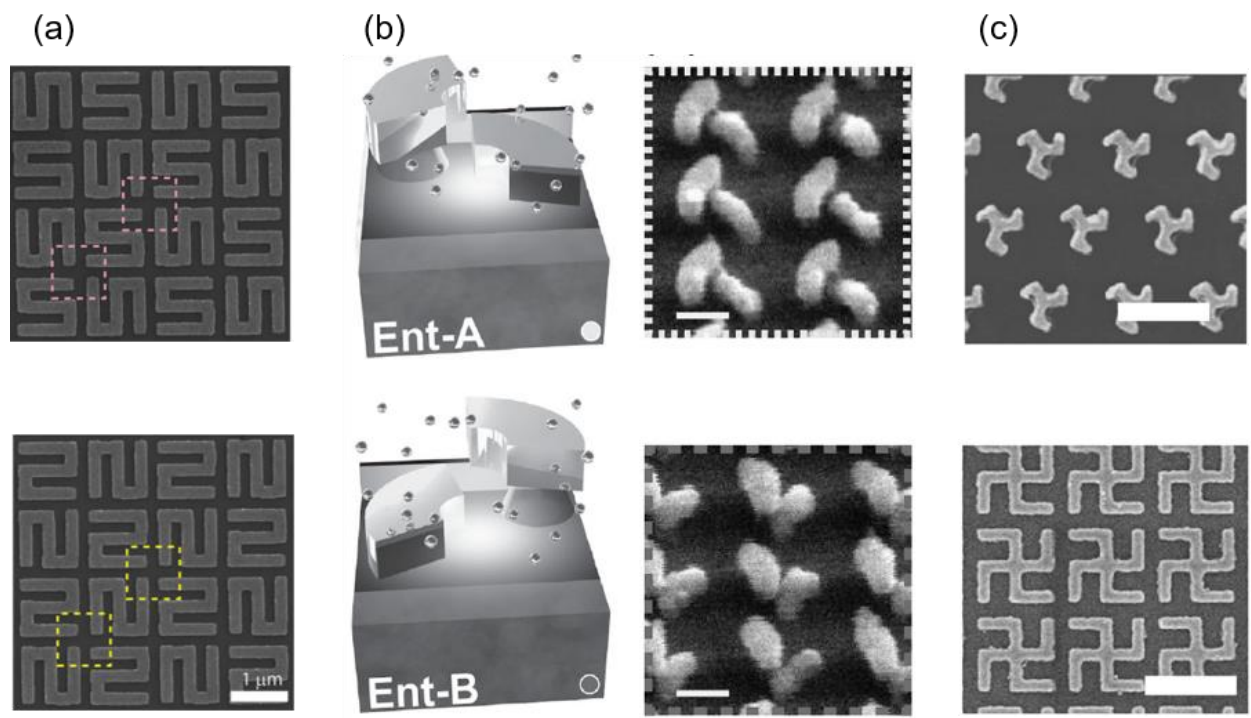

**Figure 7.** Nonlinear chiral metamaterials. Nonlinear chiroptical properties, including SHG-CD and THG-CD, are observed in various geometries, such as (a) the rotating-S,[82] (b) twisted-arcs,[92] and (c) trisceli- and gammadion-type structures.[88]

## 3. Conclusion and future prospects

To summarize, we have reviewed the recent advancements in chiral metamaterials with a designated focus on the active and nonlinear aspects. Other than more conventional methods, such as photoexcited carriers, phase-change materials, and electrical doping, we also summarized a relatively new mechanism of magneto-chiral effect for active tuning. The figure of merit for this effect is the magneto-chiral anisotropy, which has been improved dramatically lately. It would be even more fascinating to find new mechanisms or scenarios that could further increase magneto-chiral anisotropy for practical applications in chemistry and pharmaceutical fields. Another promising way to achieve active tuning is to reconfigure the structures of chiral metamaterials, for example, using mechanical methods and biochemicals. Besides directly modulating the chiral metamaterials, the modulation can also occur near to the metamaterial using liquid crystals or mechanical membranes. In the nonlinear optical regime, chiral metamaterials are particularly useful for investigating chiroptical properties. We thus also reviewed nonlinear chiral metamaterials based on second and third-order nonlinearities. Interestingly, the third-harmonic Rayleigh scattering optical activity observed recently is very user-friendly, which may find many applications in identifying nonlinear chiroptical properties in biomedical and chemical species.

One of the challenges for tuning chirality is to achieve a high modulation depth or to switch the handedness completely at strong circular dichroism or optical activity with low energy consumption. To address that challenge, the first step is to create chiral metamaterials that only respond to one selected handedness while being inert to the other. One example is to use spin preserving meta-mirrors with structured metals so that one handedness of the circularly polarized light has zero reflection.[93] With the intrinsic material loss of metals, the chirality cannot reach the maximum. For an enhanced chirality, Gorkunov et al. employed an emerging concept of bound states in the continuum with dielectrics.[94] The tailored resonant metamaterials only couple to one circular polarization of light so that the co-polarized transmission is more than 90%. Fabrication errors and defects will cause scattering loss in addition to the material loss. Spin-momentum locked topological photonics immune to scattering loss could thus further enhance the chirality. The second step is to look for novel materials and structures that efficiently complete the switching process. To that end, some promising candidates are phase change materials and two-dimensional materials.

From the device perspective, creating monolithic devices using chiral metamaterials hybridized with two-dimensional materials can dramatically reduce the device footprints, thus potentially decreasing energy consumption. For example, detecting the full-Stokes polarization of light is vital for many applications but typically demands complex and bulky optical systems. Li et al. achieved this functionality in a monolithic chiral device by integrating chiral metamaterials with graphene.[95] Besides full-Stokes polarimetry, it would be intriguing to see how hybrid monolithic devices could reach many more functionalities, such as spectral detection and tunable chiral light sources on a chip. Also, two-dimensional materials do not have to be limited to more conventional semimetals, semiconductors, or dielectrics. They can also be two-dimensional ferromagnetic, ferroelectric, and magnetic materials. Therefore, monolithic and hybrid chiral devices are promising for their contributions to valleytronics, spintronics, and chiral quantum optics.

81. Collins, J. T. et al. Chirality and chiroptical effects in metal nanostructures: Fundamentals and current trends. *Adv. Opt. Mater.* **5**, 16, 1700182 (2017).

82. Valev, V. K. et al. Nonlinear superchiral meta-surfaces: Tuning chirality and disentangling non-reciprocity at the nanoscale. *Adv. Mater.* **26**, 24, 4074–4081 (2014).

83. Valev, V. K. et al. The origin of second harmonic generation hotspots in chiral optical metamaterials. *Opt. Mater. Express*, **1**, 1, 36-45 (2011).

84. Cui, Y., Kang, L., Lan, S., Rodrigues, S. & Cai, W. Giant chiral optical response from a twisted-arc metamaterial. *Nano Lett.* **14**, 2, 1021-1025 (2014).

85. Gui, L. et al. Nonlinear born-kuhn analog for chiral plasmonics. *ACS Photonics* **6**, 12, 3306–3314 (2019).

86. Ren, M., Plum, E., Xu, J. & Zheludev, N. I. Giant nonlinear optical activity in a plasmonic metamaterial. *Nat. Commun.* **3**, 1, 1–6 (2012).

87. Rodrigues, S. P. et al. Intensity-dependent modulation of optically active signals in a chiral metamaterial. *Nat. Commun.* **8**, 1, 1-8 (2017).

88. Chen, S. et al. Giant nonlinear optical activity of achiral origin in planar metasurfaces with quadratic and cubic nonlinearities. *Adv. Mater.* **28**, 15, 2992–2999 (2016).

89. Chen, S. et al. Symmetry-selective third-harmonic generation from plasmonic metacrystals. *Phys. Rev. Lett.* **113**, 3, 033901 (2014).

90. Ohnoutek, L. et al. Optical activity in third-harmonic rayleigh scattering: A new route for measuring chirality. *Laser Photon. Rev.* **15**, 11, 2100235 (2021).

91. Ohnoutek, L. et al. Third-harmonic mie scattering from semiconductor nanohelices. *Nat. Photon.* 1–8 (2022). doi:10.1038/s41566-021-00916-6

92. Rodrigues, S. P., Lan, S., Kang, L., Cui, Y. & Cai, W. Nonlinear imaging and spectroscopy of chiral metamaterials. *Adv. Mater.* **26**, 35, 6157-6162 (2014).